\documentclass[aps,prl,floatfix,twocolumn,showpacs,10pt]{revtex4}
\usepackage{graphicx}

\begin{document}
\title{Observation of  Metastable and Stable Energy Levels of   EL2  in Semi-insulating  GaAs}

\author
{D. Kabiraj$ ^{\star}$ and Subhasis Ghosh$^{\dag} $}

\affiliation{$ ^{\star}$Nuclear Science Centre, New Delhi 110067}
\affiliation{ $ ^{\dag}$School of Physical Sciences, Jawaharlal Nehru University, New Delhi 110067}

\begin{abstract}
By using combination of detailed experimental studies, we identify the  metastable  and stable energy levels of EL2 in   semi-insulating GaAs. These results are discussed in the light of the  recently  proposed models  for stable and metastable configurations of EL2 in GaAs.   
\end{abstract}

\maketitle

  Physics of metastable  point defects is one  of the most fascinating  problems in contemporary condensed matter  physics. Over the years several metastable defects in both covalent and ionic solids have been discovered, but for technological reasons, EL2 in GaAs is  the most intensely studied, yet least understood,   metastable defect. It is believed\cite{jn99}  that bond-breaking mechanism(BBM), in which defects change lattice position is universally responsible for all the metastable defects in solids. 
The EL2 defects\cite{dr96}, responsible for semi-insulating(SI) properties of GaAs by compensating residual acceptors, can exists in two different atomic configurations. The first one is the well characterized normal configuration(EL2$^n$) and the second one is the  metastable configuration(EL2$^m$) in which all the optical, electrical and magnetic properties of EL2 disappear when SI-GaAs is illuminated with a subband gap light $\sim$1.1eV at low temperature($\leq$150K), known as phtoquenching(PQ).  EL2$^m$ is metastable because  all the properties can be recovered either by heating the sample, or by phtoexcitation at low temperature, known as photorecovery(PR). Though, a great deal of attention has already been paid to understand the physics of this defect,  but the microscopic origin of EL2 remains illusive till today.   In particular, the most important issues remained to be resolved    are,  (i)  atomic configuration of EL2 in both normal and metastable  configurations,  (ii)    driving mechanism for EL2$^n$ to EL2$^m$ transition  and,  (iii) compensation mechanism  after PQ, since free holes with concentration same as that of EL2 will be available when EL2s are in photoquenched state.  There are some efforts\cite{tb92,aa96,to93,ms95} to explain some of these issues by postulating the existence of actuator levels which trigger the metastable transition by capturing  photoexcited holes from EL2s. But,  this actuator level is characterized by the absence of any direct experimental observation.   In addition to this, interest in EL2 has been revived for two reasons, first: an increasing interest in  {\sl defect engineering} using this defect and second: the role of this defect   on the properties of low temperature grown GaAs.

Several models\cite{dr96}  based on either isolated native point defect or defect complex comprising of arsenic anitisite(As$_{Ga}$), gallium antisite(Ga$_{As}$), arsenic vacancy(V$_{Aa}$), gallium vacancy(V$_{Ga}$), arsenic interstitial(As$_i$) have been proposed, but very little is known  about the  electrical properties and  defect energy levels of EL2$^m$.  Recently, two models, for the first time, proposed  correlation between electrical activity and the defect energy level related to metastable configuration of EL2.   Fukuyama et al\cite{af03} have proposed a  BBM-based model and    shown that a three-center-complex   V$_{As}$-As$_{Ga}$-Ga$_{As}$ is the atomic origin of EL2 with specific predictions of   different defect energy levels related to EL2 in normal and metastable configurations and their role on  PQ(EL2$^n$ $\rightarrow$ EL2$^m$) and PR(EL2$^m$ $\rightarrow$ EL2$^n$). Chadi\cite{djc03} has also  proposed a  BBM-based model  and reaffirmed  that    atomistic origin of EL2 is the isolated As$_{Ga}$ which can exist in eight charge states giving rise to two energy levels for  EL2$^n$ and EL2$^m$.    It is only known that EL2 introduces a mid-gap level.  There is no  experimental investigations on the normal and metastable levels proposed in  Ref.\cite{af03} and Ref.\cite{djc03}. Identification of these levels will help immensely to resolve a long standing controversy  regarding the microscopic origin of  EL2$^n$ and  EL2$^m$ and mechanism behind the EL2$^n$ $\rightarrow$ EL2$^m$ transition.    In this letter, we report  an experimental observation of  the energy levels directly related to EL2 in normal and metastable states using different  spectroscopic techniques suitable for highly resistive materials  like,  SI-GaAs.

{\sl Experiment}. The photo-current (PC), thermally stimulated current (TSC), photo-Hall-voltage (PHV) and thermally stimulated Hall voltage (TSHV) measurements on both unirradiated and irradiated SI-GaAs samples obtained from two different sources were carried out. The observation of TSC and TSHV spectra with the signature of trap levels requires enough carriers be available for capture and finally for emission  and is achieved by the photoexcitation at low temperature prior to  heating cycle. Three  monochromatic sources, 1.16eV and 1.37eV(which cause PQ)  and 2.54eV(which does not cause PQ)\cite{aa96,vp91} were used for this purpose.   In order to perform the
TSC and TSHV measurements under identical initial
condition, following
sequence was  used,  (i) the sample was cooled to 10K in dark  and equilibrated for two hours and then
photoexcited  for PC growth and finally illumination was
terminated   at different times for obtaining the  TSC spectra  from 10K to 300K by heating the sample at the rate of 0.1K/sec,  (ii) the
sample temperature was raised to 320K and equilibrated for two
hours to avoid  any residual metastability, (iii) the sample temperature was again brought down to
10K  and step (i) was followed.   To support  the microscopic identification, 48MeV Li ions,  obtained from the Pelletron accelerator at Nuclear Science Centre, New Delhi, was used  to create and annihilate the    intrinsic point defects in SI-GaAs.     The energy  is chosen such a way that the range of the ions is larger than the thickness($\sim100\mu m$) of the sample, so that the damage of the sample due to the creation of extended defects by nuclear collision can be avoided and   the intrinsic defects are created only by electronic excitation\cite{jfz85}.

Fig.1(a) shows the PC growth at 10K under the excitation of 1.16eV and 2.54eV lights. As expected there is no PQ of EL2 in case of 2.54eV light, but strong PQ has been observed after 50sec of photoexcitation  in case of 1.16eV light and subsequently PC is saturated to a steady state value, which depends on the intensity of the photoexcitation.  Fig.1(b)   shows the comparison of  TSC spectra taken after 20sec exposure with 1.16eV and 2.54eV lights, respectively.   The most interesting feature is the completely different TSC spectra in these two cases. The peaks at 26K and   140K  are absent in TSC spectra taken with 2.54eV light and peak at 75K, 240K and 260K are absent in the TSC spectra taken with 1.16eV light. Fig.1(c) shows the comparison of TSC spectra taken after 600sec exposure with 1.16eV and 2.54eV lights.  In contrast to previous case,  TSC spectra are almost identical in this case.  Hence, there is  no difference in TSC spectra taken with 2.54eV light for different exposure times, but TSC spectra taken with 1.16eV light are completely different for different exposure times.

To study the PQ and change in conduction type during quenching, time evolution of PC and PHV growth has been studied under  the photoexcitation  with two different quenching  lights(1.16eV and 1.37eV), which are shown in Fig.2 and Fig.3. It is known\cite{af03} that SI-GaAs is very weakly n-type. The temporal evolution of PC and PHV can be divided into two regimes, regime I: a steep rise  followed by  decrease towards a minimum due to PQ of EL2(EL2$^n$ $\rightarrow$ EL2$^m$) and conduction is p-type in this regime, regime II: PC and PHV reach a minimum,  conduction type changes from p-type to n-type, followed by a   very slow  increase due to PR of EL2(EL2$^m$ $\rightarrow$ EL2$^n$) and finally conduction type returns to weakly n-type.   In case of 1.16eV light(Fig.2), PQ is over after 50sec, followed by PR which restores initial PC and conduction type after 550sec, but in case of 1.37eV light(Fig.3), it takes longer time($\sim$1500sec) for PQ and efficiency of PR is very low compared to 1.16eV light. This is due to strong dependence quantum efficiency of PQ and PR on the wavelength of photoexcitation\cite{aa96,vp91}. To investigate the evolution of TSC spectra at different stages of quenching and recovery of EL2,  light exposure was terminated at different stages of PQ and PR and the corresponding  TSC spectra show how the different energy levels evolve with quenching and recovery of EL2.  Fig.2 and Fig.3 show similar TSC spectra irrespective of wavelength of initial photoexciation. It is clear from Fig.2 and Fig.3 that energy levels at 26K and 140K   are disappearing as  EL2$^n$ $\rightarrow$ EL2$^m$, so these energy levels must be  related to metastable configuration of EL2.   After PR, these metastable levels are not recovered, instead, a new set of levels at 75K,  240K and 260K are appeared, so these energy levels are observed as EL2$^m$ $\rightarrow$ EL2$^n$.  Similar metastable TSC peak at around 140K has been previously observed\cite{zqf93,dcl97}, but not the other metastable TSC peaks.  As shown  in Ref.\cite{vp91}, we have also observed two step PQ(shown in Fig2. and Fig.3) which will be subject of our future investigation. 

Fig.4 shows the  TSC and TSHV spectra taken after 100sec exposure(during PQ) with  of 1.16eV light. By comparing the TSHV and TSC spectra, we observe, (i)  the metastable TSC peaks at 26K and 140K are present in TSHV spectra   and hence these are hole traps\cite{rk96},  (ii) the  TSC peaks at 65K, 90K  and 120K are absent, instead a dip at their positions and these should be due to electron traps\cite{rk96}, which should give rise to negative peak in TSHV spectra.    

The irradiation induced modifications in PC and TSC measurement results are shown in Fig.5, which  provide  useful information regarding the role of the levels at 26K and  140K  observed during  EL2$^n$ $\rightarrow$ EL2$^m$ transformation.  Fig.5(a) shows how efficiency of PQ is reduced compared to that in control  sample in samples  irradiated with a  fluence of 1$\times$10$^{12}$ions/cm$^2$  and  1$\times$10$^{13}$ions/cm$^2$. Corresponding TSC spectra  in Fig.5(b), clearly show that the  heights of the  metastable peaks at 26K and  140K responsible for PQ and EL2$^n$ $\rightarrow$ EL2$^m$ transition  reduced in sample irradiated with a  fluence of 1$\times$10$^{12}$ions/cm$^2$ and finally TSC peak at 26K  disappeared and the peak height of TSC peaks at 140K  reduced further in samples irradiated with a  fluence of 1$\times$10$^{13}$ions/cm$^2$. Hence, there is correlation between the reduction of efficiency of PQ i.e. EL2 concentration and reduction of the peak heights of the metastable peaks at 26K and 140K related to  EL2$^m$.

{\sl Origin of TSC/TSHV peaks in SI-GaAs when EL2s are in normal state}. The TSC spectra obtained either after photoexcitation of 2.54eV light or after photoexcitation of 1.16eV(or 1.37eV)for long duration(after PR)  are the energy levels in SI-GaAs when EL2s are in  normal state. The origin of weak TSC peak at 40K  is not known much, recently it has been shown\cite{hgs04} that it may be due to complex related to V$_{Ga}$ and Ga$_{As}$. The strong peak at 65K and 120K(as shown in Fig.1, Fig.2 and Fig.3) are    due to two energy levels of double donor V$_{As}$(+/0 and 2+/+)\cite{zqf93,dcl97}.  We have shown\cite{dk04} that the strong peak at 90K may be due to O$_{As}$. This identification is corroborated by the   annihilation of this peak under ion irradiation with 50MeV Li, as shown in Fig.5.   Based on earlier published data\cite{zqf93,dcl97,mp00} and  energetic positions, the TSC  peaks  at 150K, 200K and 215K are  attributed  to  As$_i$-V$_{As}$, Ga$_{As}$-V$_{As}$ and  As$_{Ga}$- V$_{As}$  complexes, respectively. This is all about the origin of TSC peaks when EL2s are in normal state, except the peaks at 240K and 260K, which will be discussed later.

{\sl Origin of metastable levels of EL2}. All these findings indicate that the metastable TSC peaks at 26K($\sim$0.035eV), 132K($\sim$0.27eV), which are observed during EL2$^n$ $\rightarrow$ EL2$^m$  are related to EL2$^m$, i.e. EL2 in metastable configuration, because of the (i) observation of these levels when EL2s are in metastable state, (ii) disappearance of these levels when EL2s are in normal state, (iii) gradual disappearance of these levels when the concentration of EL2 are reduced in irradiated samples, (iv) similar electrical property of all these levels, which are hole-traps,  and (v) absence of these levels in Cr-doped SI-GaAs(not shown).   These results can be interpreted by  both  the models discussed in Ref.\cite{af03} and Ref.\cite{djc03}. But, to put our experimental results on more rigorous basis we now  discuss the three-center-complex model based on the proposal of V$_{As}$-As$_{Ga}$-Ga$_{As}$, as atomistic origin of EL2. According to this model, (i) $As_{Ga}-V_{As}$ pair is responsible for metastability while $Ga_{As}$ controls the transition between the normal and metastable state, (ii) the Columbic interaction between $Ga_{As}^-$ and $As_{Ga}^+$ pins the As atom at its Ga-site. PQ starts with the  photoionization of hole from $As_{Ga}^+$($As_{Ga}^+ \rightarrow As_{Ga}^0 + h$) and subsequent capture of   hole by  $Ga_{As}^-$, which is supported by the p-type conductivity during PQ, as shown in Fig.2 and Fig.3. This results the neutralization of $Ga_{As}^-$ and $As_{Ga}^+$ to $Ga_{As}^0$ and $As_{Ga}^0$, respectively, leading to (i) switching off the  Coulombic attraction  and  breakage of the bond  between  $As_{Ga}$ and $Ga_{As}$ and (ii) As atom moves towards the V$_{As}$  due to BBM until the formation of metastable complex.  Movement of As atom from As$_{Ga}$ site towards the interstitial site gives rise to V$_{Ga}$ and As$_i$. The role of  As$_{Ga}$ and movement of As$_i$ by few angstroms on the EL2$^n$ $\rightarrow$ EL2$^m$ transition has also been established theoretically\cite{djc03,djc88}. Hence, as EL2$^n$ $\rightarrow$ EL2$^m$, these metastable point defects(V$_{Ga}$ and As$_i$) should give rise to TSC peaks, which should be  hole traps. The metastable defect energy levels at 0.035eV and 0.27eV from valence band minimum(VBM)     are attributed to  As$_i$ and $V_{Ga}$, respectively. Similar ionization energies of V$_{Ga}$ and As$_i$ have been  predicted by calculating local atomic structure with  lattice relaxation\cite{hs95,jts02}. The existence of V$_{Ga}$ in the metastable configuration of EL2$^m$ has been argued  by positron annihilation experiment\cite{rk90}. 

The recovery of EL2 according to this three-center-complex model would be $[V_{As}-(As)-V_{Ga}]-Ga_{As}^0 \rightarrow [V_{As}^+-As_{Ga}^+]-{Ga_{As}^-} + e$, and the excess electrons in the conduction band during this process is consistent with experimentally observed(Fig.2 and Fig.3) conduction type conversion from p-type during PQ to n-type during recovery of EL2\cite{af03}. According to this model, three processes: (1) ionization of deep donor $V_{As}^0$, which is supported by the enhancement of TSC peaks at 65K and 120K during PR, as shown in Fig.2 and 3,  (2) ionization of deep acceptor $Ga_{As}^0$, which should result a new metastable peak during PR and is attributed to the TSC peak at 75K($\sim$0.13eV), similar ionization energy of Ga$_{As}$(0/-) has also been predicted theoretically\cite{mjp89,sbz90},  and (3) movement of As atom towards the antisite position resulting annihilation of As$_i$ and V$_{Ga}$, which is supported by the disappearance of TSC peaks at 26K the 140K during PR. 

{\sl Origin of stable  levels of EL2}. As the recovery process proceeds with prolonged  photoexcitation with sub band gap light, we observe the disappearance of  metastable levels related to EL2 and  appearance of new   peaks at  240K($\sim$0.49eV)  and 260K($\sim$0.56eV). These peaks are always observed when TSC is taken with above band gap light. These levels are  related to the normal state of EL2. Based on their energetic positions, these levels are attributed to  double acceptor Ga$_{As}$(-/2-) and double donor As$_{Ga}$(2+/+)\cite{hs95,jts02}.  To have EL2s in normal state,  Coulombic interaction  is required between  Ga$_{As}$ and As$_{Ga}$, so they should be either in Ga$_{As}^-$ and As$_{As}^+$ or Ga$_{As}^{--}$ and As$_{As}^{++}$ states.  Recently, similar ionization energy, 0.47eV from VBM, which should be equivalent to TSC peak at 240K,  has been observed by excitation photocapacitance  method\cite{yo05}.   	 
Lastly, we would like raise a issue regarding the observation of these stable levels only 
when EL2s are always in normal state i.e. never photoquenched at low temperature. Once the temperature is raised beyond 150K, they should be observed always. Is  the  thermal recovery not complete beyond 150K ? Self promoted behavior of EL2 and temperature dependent {\sl incubation time}  have been observed\cite{af03} during recovery and explained by  some kind of correlation among EL2s, but more experimental and theoretical studies are required for the ascertainment of this speculation. It has been shown that  correlated defects are required for consistent description of experimental results  in highly compensated Ge\cite{kmi96} and magnetic semiconductor (Ga,Mn)As\cite{ct02}.

In conclusion, we  have identified  the defect energy levels related to normal and metastable configurations of EL2 in SI-GaAs by using different spectroscopic techniques  and PC growth during PQ and PR under different initial conditions. A three center model for microscopic structure of EL2 has been discussed in the context of the origin of these metastable and stable levels and their role on PQ and PR of EL2 in SI-GaAs.

\section*{Figure Captions}

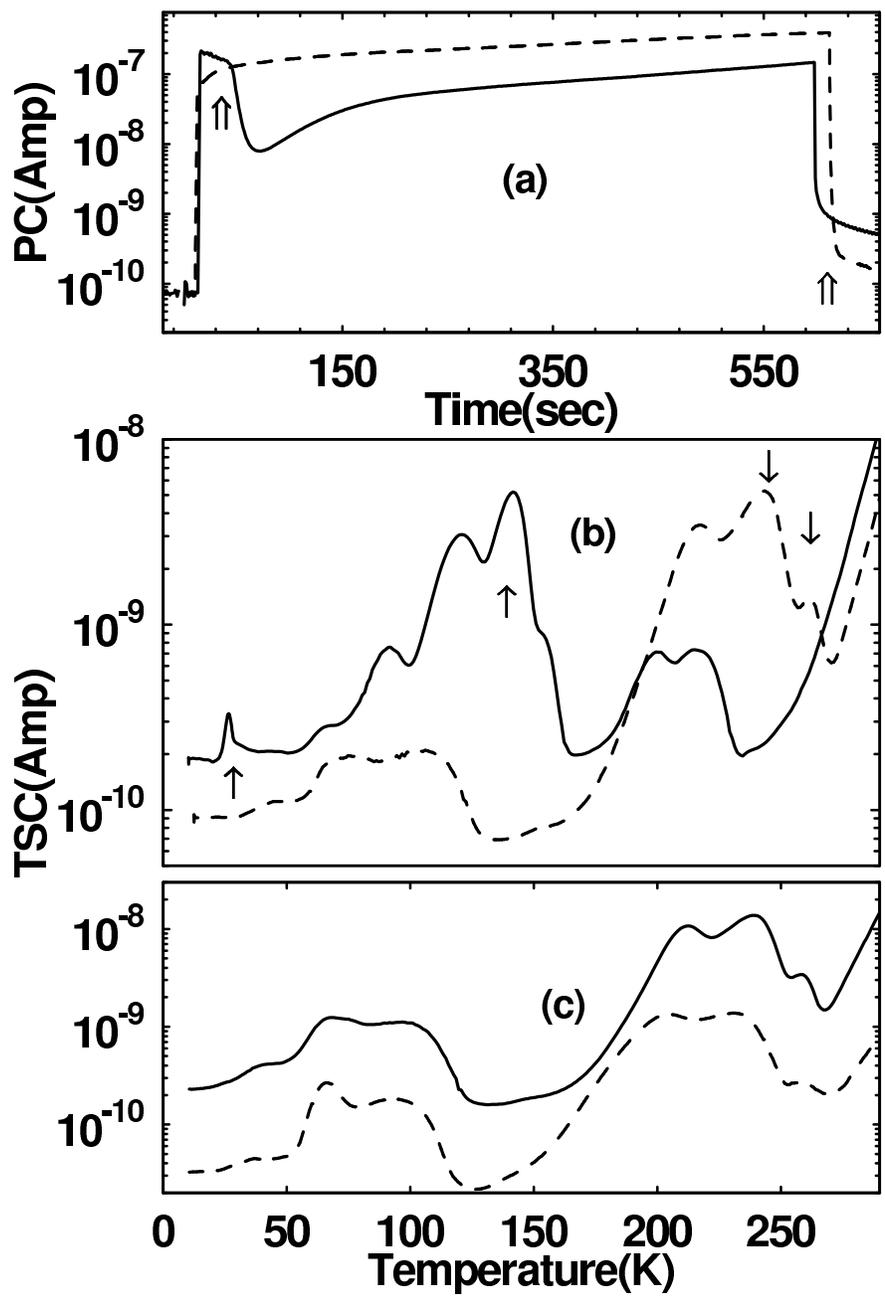
\noindent Figure 1. (a) PC growth in SI-GaAs with 1.16eV light(solid line) and 2.54eV light(dashed line). Photoexcitation terminated at different times(shown by $\Uparrow$ in (a)) and corresponding TSC spectra taken  with 1.16eV and 2.54eV lights  for (b) 20sec and (c) 500sec exposure at 10K. Light was switched on at t=10sec.

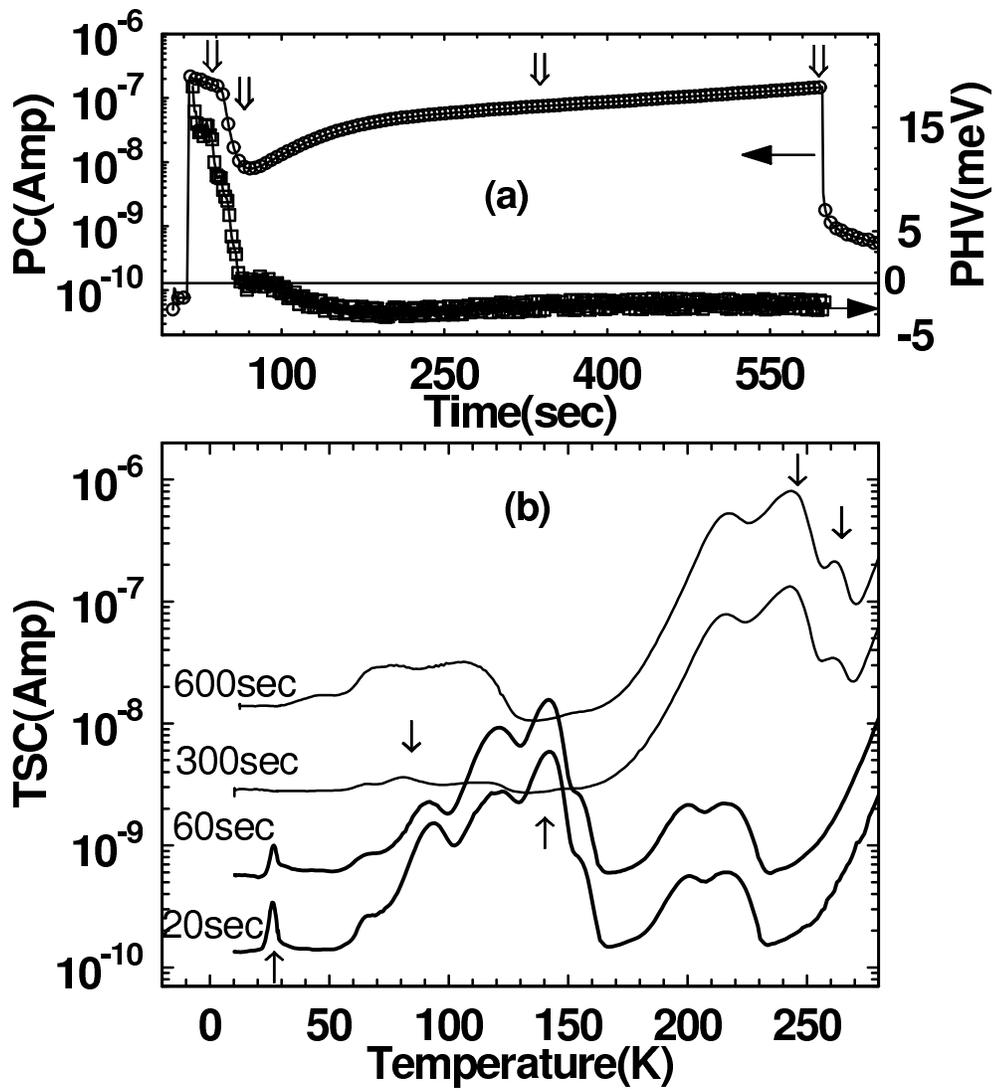
\noindent Figure 2. (a) Comparison of temporal evolution of PC growth and photo Hall voltage(PHV) under the photoexcitation of 1.16eV light. Light was switched on at t=10sec. (b) Evolution of TSC spectra under the different dose of initial photoexcitation(shown by $\Uparrow$ in (a)).  TSC  peaks observed during PQ(EL2$^n$ $\rightarrow$ EL2$^m$) and PR(EL2$^m$ $\rightarrow$ EL2$^n$) are indicated  by $\uparrow$ and  $\downarrow$, respectively. TSC spectra(except the bottom one) are shifted up for clarity.

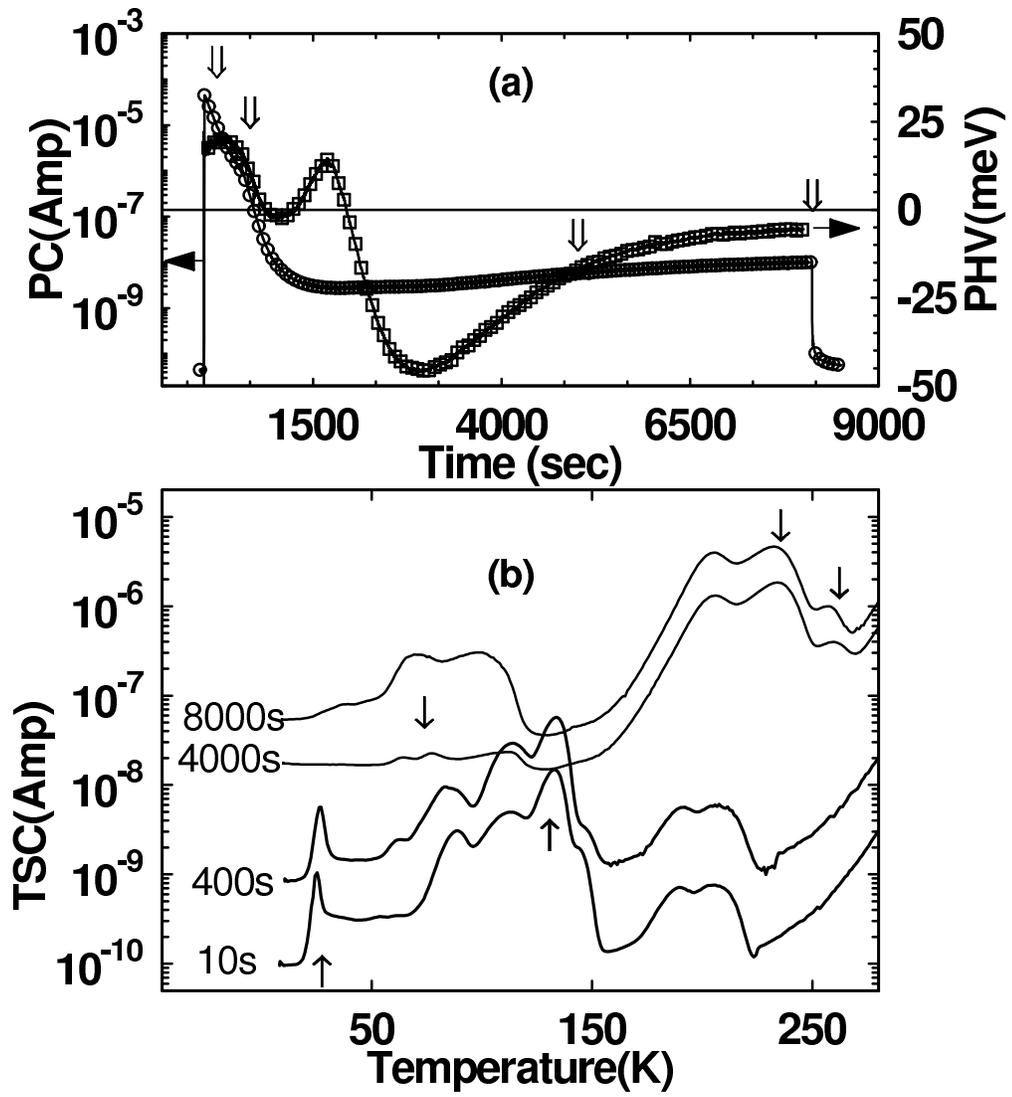
\noindent Figure 3. (a) Comparison of temporal evolution of PC growth and photo Hall voltage(PHV) under the photoexcitation of 1.37eV light. Light was switched on at t=10sec. (b) Evolution of TSC spectra under the different dose of initial photoexcitation(shown by $\Uparrow$ in (a)).  TSC  peaks observed during PQ(EL2$^n$ $\rightarrow$ EL2$^m$) and PR(EL2$^m$ $\rightarrow$ EL2$^n$) are indicated  by $\uparrow$ and  $\downarrow$, respectively. TSC spectra(except the bottom one) are shifted up for clarity.

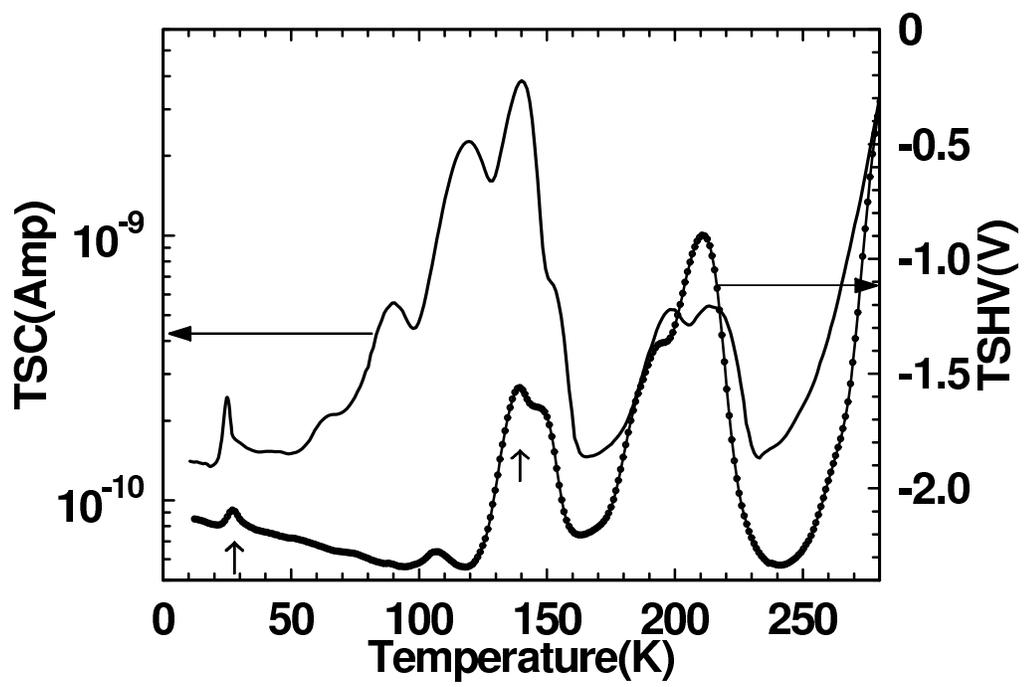
\noindent Figure 4. Comparison of TSC(solid line) and TSHV(dotted line) spectra during PQ i.e. when EL2s are in metastable states. The peaks related to EL2$^m$ are indicated by arrow($\uparrow$).

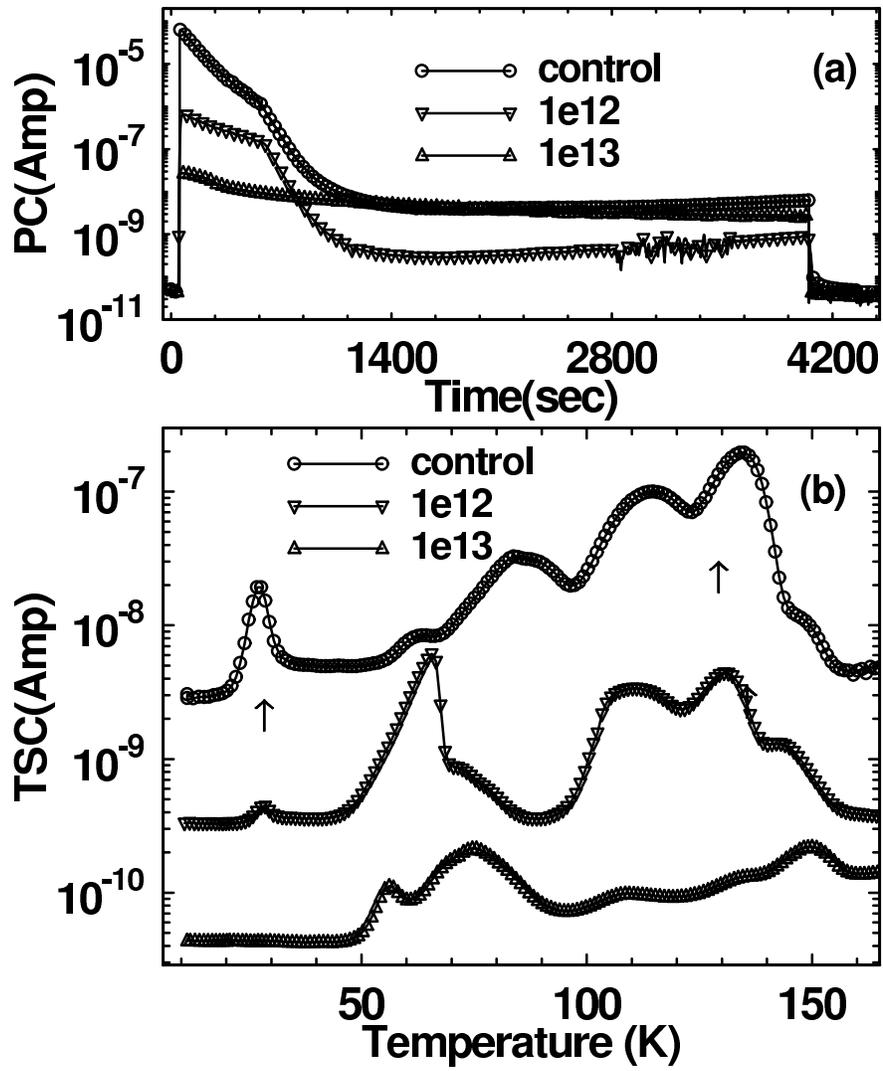
\noindent Figure 5. (a) PC growth in irradiated SI-GaAs samples with 1.37eV light. Light was switched on at t=10sec. (b) Evolution of TSC spectra in irradiated SI-GaAs samples, with 20sec initial photoexposure.

\end{document}